# Bismuth doping induced enhancement of the spin–orbit coupling strength in the prototype dilute ferromagnetic semiconductor (Ga,Mn)As: A review

Tadeusz Wosinski

*Institute of Physics, Polish Academy of Sciences, Aleja Lotnikow 32/46, PL-02668 Warsaw, Poland*
E-mail: wosin@ifpan.edu.pl

**Abstract**

Extensive studies on the impact of bismuth incorporation into the (Ga,Mn)As prototype dilute ferromagnetic semiconductor (DFS) on its structural, magnetic and magnetotransport properties are summarized in this review. Thin epitaxial layers of the quaternary (Ga,Mn)(Bi,As) compound, containing up to 1% Bi and 6% Mn atoms, and the reference ternary (Ga,Mn)As compound, have been grown under either a compressive or tensile biaxial misfit strain by the low-temperature molecular-beam epitaxy technique with precisely optimized growth conditions. The high-resolution X-ray diffractometry (HR-XRD) measurements and transmission electron microscopy (TEM) imaging of cross-sections across the sample interfaces have evidenced for high structural perfection of the DFS layers and sharp interfaces with the substrate. An addition of bismuth into the layers causes a small decrease in their ferromagnetic Curie temperature and a distinct increase in the coercive fields, as revealed by the superconducting quantum interference device (SQUID) magnetometry investigations, which also demonstrate a strong effect of biaxial misfit strain in the layers on their crystalline magnetic anisotropy. Most of all, the incorporation of a small atomic fraction of heavy Bi atoms, substituting As atoms in the layer, predominantly enhances the spin–orbit coupling strength in its valence band, considerably affecting electromagnetic properties of the layers. Investigations of magnetotransport properties of the DFS layers, performed on micro-Hall-bars prepared from the layers using electron-beam lithography patterning and chemical etching, reveal, as a result of Bi addition to the layers, significantly enhanced magnitudes of magnetoresistance, anomalous and planar Hall effects as well as the spin–orbit torque effect. The latter effect is of special interest for applications to the next generation non-volatile data storage and logic spintronic devices, utilizing electrically controlled magnetization reversal.

**Keywords:** dilute ferromagnetic semiconductors; molecular-beam epitaxy; magneto-crystalline anisotropy; magnetotransport phenomena; spin–orbit coupling; spintronics

**Introduction**

The ternary III-V semiconductor compound (Ga,Mn)As, successfully grown over two decades ago [1,2], has been recognized as the prototype dilute ferromagnetic semiconductor (DFS) contributing to develop new spintronic functionalities for future electronics [3,4]. Nonequilibrium growth conditions of low-temperature (150–250°C) molecular-beam epitaxy (LT-MBE), used to overcome the low equilibrium solubility of transition metal elements in GaAs, enable to grow homogeneous layers of (Ga,Mn)As with the content up to above 10% of Mn atoms [5–7]. Mn ions, substituting Ga ones at the cation sites in GaAs crystal lattice, act both as magnetic moments, with half-filled $3d$ shell and S = 5/2 spin moment in the $Mn^{2+}$ charge state, and as acceptor dopants ensuring $p$-type conductivity of the compound. Below the ferromagnetic Curie temperature, $T_C$, the Mn moments become aligned driven by their interactions with spin-polarized holes. The effect has been originally interpreted within the frames of the $p$-$d$ Zener model of hole-mediated ferromagnetism, proposed by Dietl et al. [8,9], which assumes the exchange coupling between the localized Mn spins and the valence-band holes. The alternative model of double-exchange mechanism of ferromagnetism, which also accounts for many features of (Ga,Mn)As layers, assumes that the itinerant holes, mediating ferromagnetic coupling of the Mn moments, reside in the Mn-induced impurity band formed above the GaAs valence band edge. Though, several various scenarios of the impurity band formation have been proposed within this model; see e.g. [10–14].

In order to differentiate experimentally between the valence-band and impurity-band origin of holes, mediated ferromagnetism in (Ga,Mn)As, we have performed a series of investigations, mainly by means of modulation photoreflectance spectroscopy, for (Ga,Mn)As layers with a broad range of 0.001% to 6% of Mn contents [15–17]. Our results of the optical energy gap evolution with increasing Mn content revealed a distinct modification of the GaAs valence band, and the Fermi level location below the valence band edge, at the Mn content of about 0.3% already, i.e., much lower than that (about 1% Mn content) required for the onset of ferromagnetic phase in (Ga,Mn)As. Those results point to the valence-band origin of holes responsible for long range ferromagnetic ordering in (Ga,Mn)As, thus supporting the $p$-$d$ Zener model. This model also properly accounts for quite a complicated crystalline magnetic anisotropy in (Ga,Mn)As layers.

Generally, the layers grown under tensile biaxial strain resulted from the lattice mismatch, like, e.g., pseudomorfically grown on the (In,Ga)As buffer with the lattice parameter larger than that of (Ga,Mn)As, display perpendicular magnetic anisotropy with the easy magnetization axis along the growth direction. In contrast, (Ga,Mn)As layers pseudomorfically grown on (001) GaAs substrates, with the lattice parameter smaller than that of (Ga,Mn)As, are grown under compressive biaxial misfit strain and demonstrate in-plane magnetization with two equivalent easy axes along two in-plane ⟨100⟩ crystallographic directions (cubic anisotropy) at low temperatures. Such behaviour reflects the anisotropic properties of the top of GaAs valence band, in agreement with predictions of the *p-d* Zener model [18]. In addition, the misfit strain relaxation in narrow stripes, lithographically patterned from thin compressively strained (Ga,Mn)As layers, results in patterning-induced (shape) magnetic anisotropy, aligning the magnetization vector along the stripes [19–21].

Strong magneto-crystalline anisotropy of (Ga,Mn)As layers and the associated very efficient planar Hall effect (PHE) have been proposed to utilize in construction of novel non-volatile memory devices with magnetic recording and electrical reading, based on a single (Ga,Mn)As layer [22–24]. On the other hand, we have suggested a construction of memory cells employing the resistance of magnetic domain walls in nanostructures of specific geometries, fabricated from a thin (Ga,Mn)As layer by means of electron-beam lithography patterning and chemical etching, and involving a competition between various types of magnetic anisotropies of the nanostructures: the cubic crystalline anisotropy of the layer and the patterning-induced one [25].

Moreover, the first experimental demonstration of switching the magnetization direction between two easy axes, driven by the current-induced spin–orbit torque (SOT) mechanism, has been performed on a (Ga,Mn)As layer [26]. The SOT mechanism of magnetization manipulation arises due to the spin–orbit coupling (SOC) phenomenon, which originates from relativistic interaction between the charge carrier's spin and its angular momentum [27,28]. In conducting materials, like (Ga,Mn)As of zinc-blende crystal structure with broken inversion symmetry, SOC results predominantly from the Dresselhaus effect [27]. Misfit strain in the (Ga,Mn)As layer, decreasing its crystal symmetry, causes additional spin–orbit interaction of the Dresselhouse type [29]. Electric current flowing through the crystal generates, as a result of spin–orbit interaction, an additional spin polarization of charge carriers, proportional to their wavevector, which exerts a torque on localized magnetic moments in the crystal, giving rise to the SOT mechanism [30,31]. This mechanism is of topical interest for the next-generation, energy efficient, non-volatile data storage and logic applications, as it has several advantages, like lower energy consumption and faster device operation, over the currently utilized spin-transfer torque mechanism [31–34].

Aiming at enhance the strength of SOC in the (Ga,Mn)As DFS we have grown epitaxial layers of the quaternary (Ga,Mn)(Bi,As) compound containing a small atomic fraction, 0.3% to 1%, of bismuth, the heaviest group-V element in the periodic table [35,36]. Earlier investigations established that incorporation of heavy Bi atoms, replacing As atoms at the anion sites in GaAs crystal, brought about a relativistic correction to the GaAs valence band structure and effectively enhanced the SOC strength in the ternary Ga(Bi,As) compound [37–39]. Large differences in the sizes and electronegativities between Bi and As atoms, causing the weak Bi–Ga bonding energy and large miscibility gap require highly nonequilibrium growth conditions of low-temperature molecular-beam epitaxy at the temperatures in the range of 200–400°C, which are significantly lower than the optimal one, of about 580°C, for the MBE growth of GaAs layers [40,41]. The low MBE growth temperature enhances the solubility of Bi in GaAs but, on the other hand, it results in the formation of structural defects, mainly $As_{Ga}$ and $Bi_{Ga}$ antisites and Bi complexes [42,43]. Various post-growth annealing treatments have been proposed to improve the electronic and optical properties of the as-grown layers [40,44]. Even more challenging becomes the growth of high-quality quaternary dilute bismides, like, e.g., Ga(As,P,Bi) and (In,Ga)(As,Bi), offering a possibility to engineer their lattice parameters and band structures to the requirements of specific optical and electronic device applications [45,46].

In the present review article we summarize our recent investigations on the impact of Bi incorporation into (Ga,Mn)As layers on their properties and the experimental evidence of phenomena arising due to enhanced SOC strength in the (Ga,Mn)(Bi,As) DFS. The properties of (Ga,Mn)(Bi,As) layers are compared with those of the reference (Ga,Mn)As layers, with the same Mn content, achieving the state-of-the-art structural and magnetic perfection and reaching the relatively high $T_C$ of 145 K for the layer with 6% Mn content [47].

**Experimental**

We have investigated a series of (Ga,Mn)(Bi,As) layers with 6% Mn and 1% Bi contents and various thicknesses, of 10, 15 and 50 nm, and, in addition, the 100-nm thick layer with 4% Mn and 0.3% Bi contents. The layers were grown by the LT-MBE technique either on semi-insulating (001)-oriented GaAs substrate or on that substrate covered with a strain-relaxed, 0.63-μm thick, $In_{0.2}Ga_{0.8}As$ buffer layer. For each (Ga,Mn)(Bi,As) layer the reference (Ga,Mn)As layer, with the same thickness and Mn content, was grown. During the growth, the $As_2$ to (Ga+Mn) flux ratio was adjusted close to the stoichiometric one and the substrate temperature, in the range of 210–230°C, was carefully selected to minimize the concentrations of arsenic antisite and interstitial Mn defects in the layers [17,48]. In-situ reflection high-energy electron diffraction (RHEED) was applied to monitor the two-dimensional growth mode and to calibrate Mn content in the layers and their thicknesses [48].

After the growth, the samples were subjected to a long-term annealing treatment, carried out in air at 180°C for a period of 80 h in the case of 100-nm thick layers and 50 h in the case of thinner layers. Such a long-term annealing at temperatures below the growth temperature evidently improves the magnetic and transport properties of the (Ga,Mn)As layers, caused mainly by out-diffusion of Mn interstitials, which contribute to the decrease in magnetic moment and hole concentration in the as-grown layers [49,50]. Similar annealing-induced improvement of magnetic properties and increase in the hole concentration have also been verified for our (Ga,Mn)(Bi,As) layers [35,36]. On the other hand, such annealing at temperatures even up to 600°C does not result in out-diffusion of Bi atoms, as evidenced from the Rutherford backscattering spectrometry experiments for Ga(Bi,As) layers [51]. Composition of the annealed DFS layers as well as of the buffer layers and the in-depth homogeneity of their compositions were confirmed by means of the secondary-ion mass spectrometry (SIMS) measurements [47].

Structural perfection of the DFS layers and quality of their interfaces have been thoroughly examined through the high-resolution X-ray diffractometry (HR-XRD) measurements and transmission electron microscopy (TEM) imaging of cross-sections across the sample interfaces. Magnetic properties of the layers have been investigated with the superconducting quantum interference device (SQUID) magnetometry through both temperature- and magnetic-field-dependent measurements, under the field applied along all the main crystallographic directions in the layers. In addition, a set of DFS layers has been subjected to muon spin relaxation (μSR) spectroscopy at the Paul Scherrer Institute in Villigen, Switzerland, employing the low-energy μSR setup at the μE4 surface muon beam line [52], which enables spatially-resolved (on nm scale) magnetic characterization of thin layers of materials by using a beam of fully spin-polarized positive muons.

Magnetotransport properties of the DFS layers have been investigated employing micro-Hall-bars prepared from the layers by means of electron-beam lithography patterning and chemical etching. The Hall-bars, supplied with electrical Ohmic contacts made by indium soldering onto the layers, have been mounted in a helium cryostat equipped with superconducting electromagnet for measurements at temperatures down to 1.5 K, in magnetic field up to ±13.5 T. Four-probe longitudinal resistance and Hall resistance of the Hall-bars have been measured using either the low-frequency ac lock-in technique or dc Keithley Delta mode one. Moreover, valence band modifications in the (Ga,Mn)As layer caused by incorporation of a small atomic fraction of Bi atoms have been inspected by means of modulation photoreflectance spectroscopy, spectroscopic ellipsometry and hard X-ray angle-resolved photoemission spectroscopy (HARPES). HARPES experiments have been carried out at beamline P22 of the Synchrotron source Petra III (DESY, Hamburg, Germany) using the time-of-flight momentum microscope end station [53].

**Results and Discussion**

**Structural Characterization**

The HR-XRD diffraction spectra for representative (Ga,Mn)(Bi,As) and (Ga,Mn)As layers of 50 nm thickness grown on GaAs substrate are shown in Fig. 1 [54]. The main figure presents the symmetrical 004 Bragg reflections and the inset shows the reciprocal lattice map (RLM) for the (Ga,Mn)As/GaAs sample, obtained for the asymmetrical –2–24 Bragg reflection. Here, the vertical axis $q_z$ corresponds to the component of the reciprocal lattice vector perpendicular to the layer surface, i.e. along the [001] crystallographic direction, and the horizontal axis $q_x$ corresponds to the vector component, lying in the diffraction plane, parallel to the surface, along the in-plane [–1–10] direction; cf. Ref. [55]. Both the axes are in the reciprocal lattice units (rlu) defined as $2\pi/d_{hkl}$, where $d_{hkl}$ is the lattice spacing of corresponding crystallographic planes. Similar RLM was obtained for the (Ga,Mn)(Bi,As)/GaAs sample. The nodes corresponding to the DFS layer and GaAs substrate are vertically aligned, indicating that their in-plane lattice parameters are the same, while the out-of-plane lattice parameter of the layer turns out to be distinctly larger than that of the substrate. This proves the pseudomorphic growth of the layer under an in-plane compressive biaxial misfit strain as a result of elastic relaxation.

Consequently, in the 004 diffraction spectra, shown in the main figure, the broad peaks corresponding to reflections from the DFS layers, occur at lower diffraction angles than the narrow ones corresponding to the GaAs substrate. Addition of 1% Bi to the (Ga,Mn)As layer causes a distinct increase in its lattice parameter perpendicular to the layer plane resulting from the increase in biaxial compressive strain. The in-plane misfit strain values of the layers, calculated from the angular positions of their 004 Bragg reflections, are 0.27% and 0.46% for the (Ga,Mn)As and (Ga,Mn)(Bi,As) layers, respectively [54]. Clear X-ray interference fringes appearing around the layer-related peaks evidence for homogeneous layer compositions and smooth interfaces with the substrate, confirmed by the high-resolution TEM imaging of cross-sections of the samples [54].

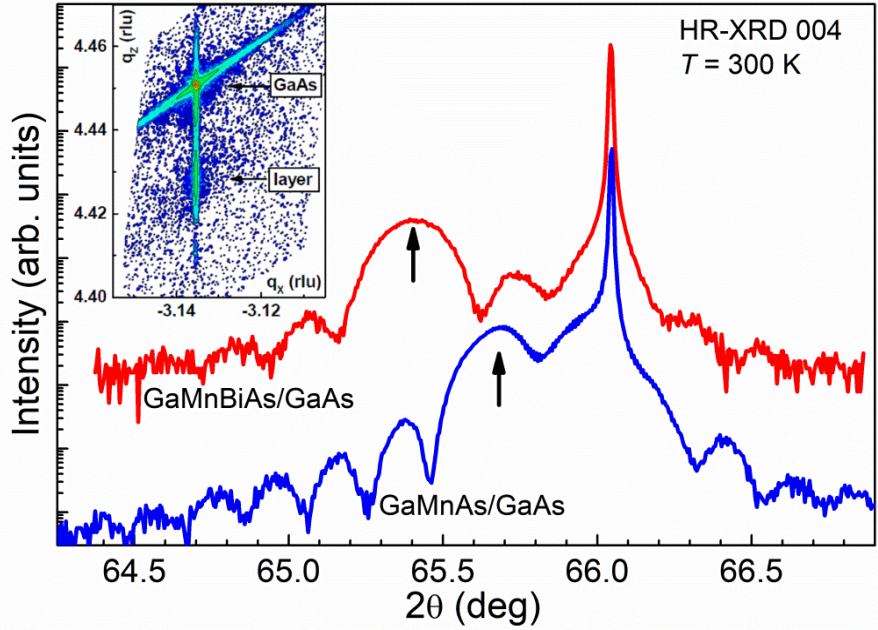

**Fig. 1** High-resolution X-ray diffraction spectra (2θ/ω scans) for the 004 Bragg reflection measured for the 50 nm thick (Ga,Mn)(Bi,As) layer with 6% Mn and 1% Bi contents, and the reference (Ga,Mn)As layer, epitaxially grown on (001) GaAs substrate. The narrow peaks correspond to the GaAs substrate and the broader ones at lower diffraction angles, indicated by the vertical arrows, are reflections from the DFS layers. The spectra are vertically offset for clarity. The reciprocal lattice map of (Ga,Mn)As/GaAs sample for the –2–24 Bragg reflection is shown in the inset. After [54].

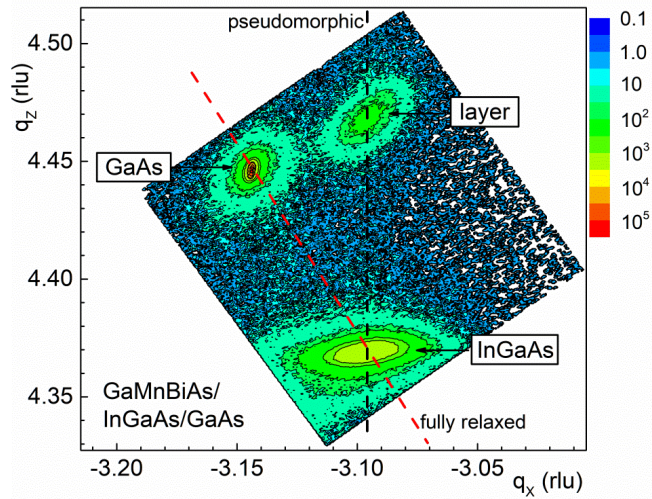

**Fig. 2** Reciprocal lattice map (RLM) of the (Ga,Mn)(Bi,As)/(In,Ga)As/GaAs heterostructure for the –2–24 Bragg reflection, where the horizontal and vertical axes are along the in-plane [–1–10] and out-of-plane [001] crystallographic directions, respectively. The diffraction nodes for the GaAs substrate, (In,Ga)As buffer and DFS layer are marked. The vertical and diagonal dashed lines correspond to the RLM node positions for pseudomorphic and fully relaxed layers, respectively. After [54].

The –2–24 reciprocal lattice map for 50 nm thick (Ga,Mn)(Bi,As) layer with 6% Mn and 1% Bi contents grown on (In,Ga)As buffer layer is presented in Fig. 2, showing the diffraction nodes for GaAs substrate, (In,Ga)As buffer and DFS layer. The vertical and diagonal dashed lines denote the RLM node positions for pseudomorphic (fully strained) and fully relaxed layers, respectively. Similar RLM was also recorded for the reference (Ga,Mn)As/(In,Ga)As/GaAs heterostructure [54]. For both heterostructures the thick (In,Ga)As buffer layers were fully plastically relaxed, while the thin DFS layers were pseudomorphically grown on the buffer under tensile biaxial misfit strain. Concluding, the HR-XRD results have shown that all the investigated DFS layers, grown either on GaAs substrate or (In,Ga)As buffer, were fully strained under either the compressive or tensile biaxial strain, respectively, indicating that their thicknesses were below the critical thickness for plastic relaxation.

High structural perfection of both the (Ga,Mn)(Bi,As) and (Ga,Mn)As layers grown on the (In,Ga)As buffers and their sharp interfaces have also been confirmed in our cross-sectional TEM analysis of the samples presented in Ref. [54]. High-resolution TEM images demonstrate perfect zinc-blend crystal structure in the layers and only few structural defects, such as threading dislocations originated at the strongly mismatched (In,Ga)As/GaAs interfaces.

**Magnetic Properties**

SQUID magnetometry measurements of the (Ga,Mn)(Bi,As) layers evidenced for qualitatively the same magneto-crystalline anisotropy as in the reference (Ga,Mn)As layers. The layers grown under tensile biaxial strain displayed the easy magnetization axis oriented along the [001] growth direction, see e.g. [47]. On the other hand, the layers grown under compressive biaxial strain displayed the in-plane magnetization. Typical temperature dependence of magnetization recorded for the layers grown on GaAs is presented in Fig. 3 [56]. At the lowest temperatures both the layers display the easy magnetization axes along two in-plane ⟨100⟩ crystallographic directions confirmed by the largest magnitude of magnetization. However, at temperatures of above about 20 K the largest magnitude of magnetization becomes along the [–110] direction indicating reorientation of the easy magnetization axis. At temperatures approaching the Curie temperature the magnetizations along all three main in-plane directions become comparable decreasing to zero at $T_C$. Such temperature dependence of magnetization is characteristic of compressively strained (Ga,Mn)As layers with sufficiently high hole concentration; cf. [56,57]. Bi incorporation into the (Ga,Mn)As layers results in slightly lower magnitudes of magnetization in the whole temperature range of ferromagnetic ordering, as measured for all three main in-plane directions, and lower, by about 15%, values of the ferromagnetic Curie temperatures, which was also confirmed for other investigated sets of (Ga,Mn)(Bi,As) and reference (Ga,Mn)As layers; cf. Refs. [35,36,47,58]. Such reduction in the Curie temperature, caused by addition of heavy Bi atoms, thus giving rise to increase in the spin–orbit coupling strength in (Ga,Mn)As layers, is in accordance with conclusions of the *p-d* Zener model of hole-mediated ferromagnetism, which predicts $T_C$ decrease as a result of increase in SOC strength in the valence band of *p*-type magnetic semiconductors [8,9].

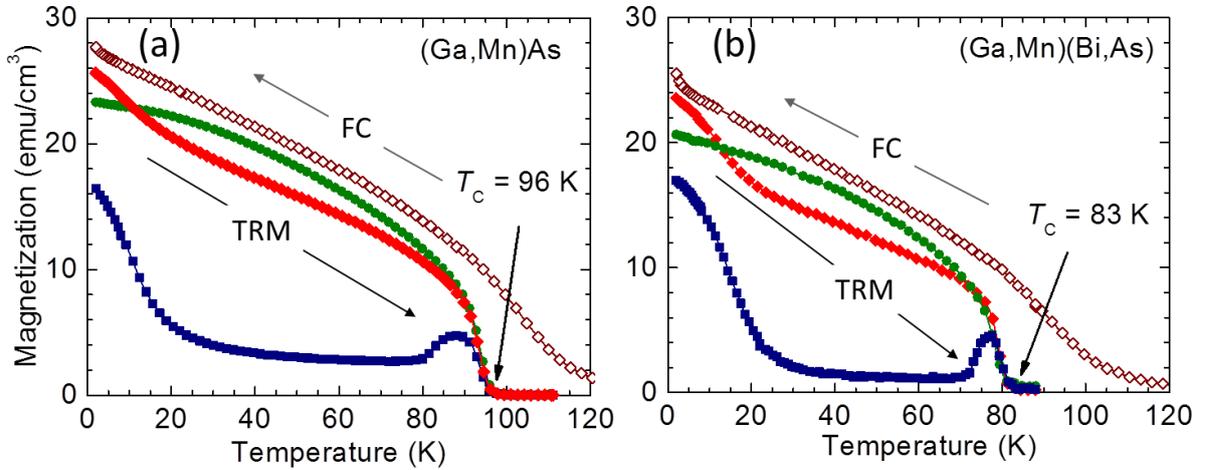

**Fig. 3** Temperature dependent magnetization of 10 nm thick (Ga,Mn)As (**a**) and (Ga,Mn)(Bi,As) (**b**) layers, with 6% Mn and 1% Bi contents, grown on GaAs. On both panels open diamonds represent the magnetization recorded during field cooling (FC) in $\mu_0 H = 0.1$ T, with magnetic field *H* applied along the [100] in-plane direction. Solid symbols mark the thermoremnant magnetization (TRM), measured upon warming the samples in the absence of *H*, right after FC, along the in-plane crystallographic directions: [100] – red diamonds, [–110] – green bullets, and [110] – blue squares. The magnitudes of the Curie temperatures, $T_C$, are indicated by arrows. After [56].

The magnetization hysteresis loops recorded at magnetic field along the main in-plane crystallographic directions at the temperature 4 K, shown in Fig. 4, confirm qualitatively the same magneto-crystalline anisotropy in both the investigated layers. Characteristic difference in magnetic properties along the two in-plane ⟨110⟩ directions (uniaxial anisotropy), with the [–110] direction being magnetically easier than the perpendicular [110] one, was explained by *ab initio* calculations [59], which predicted the preferred formation of Mn dimmers along the [–110] crystallographic direction at the (001) surface during the epitaxial growth of (Ga,Mn)As layers. The enhanced SOC strength in the Bi-contained layer results in a distinct increase in the layer coercive fields by a factor of about 1.5 for the easier [100] and [–110] crystallographic directions and of above 2 for the hard [110] one. Similarly, the tensile-strained (Ga,Mn)(Bi,As) layers displayed significantly larger coercive field along the [001] growth direction (easy magnetization axis) than that of the reference layers without Bi content [47].

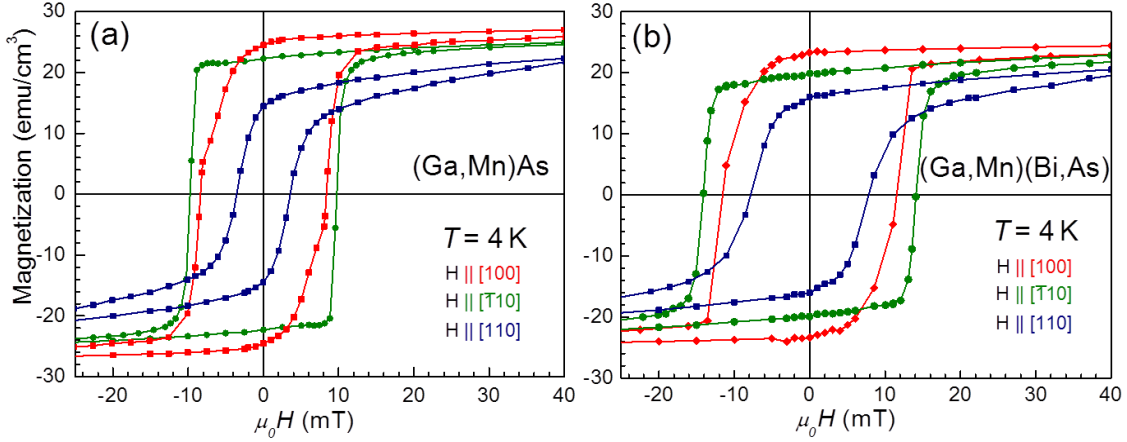

**Fig. 4** Magnetic field dependence of the in-plane magnetization (magnetization hysteresis loops), of the same (Ga,Mn)As (**a**) and (Ga,Mn)(As,Bi) (**b**) layers as in Fig. 3, measured at temperature $T = 4$ K. Magnetic field $H$ has been applied along three main in-plane directions, as described in the panels. After [56].

Spatially-resolved muon spin relaxation spectroscopy measurements have been carried out on the large area (about 1 cm$^2$) 50-nm thick (Ga,Mn)(Bi,As), with 6% Mn and 1% Bi contents, and similar reference (Ga,Mn)As layers, grown both under compressive and tensile misfit strain. The μSR spectroscopy results clearly demonstrate that the ferromagnetic phase develops uniformly in the whole volume of the (Ga,Mn)(Bi,As) layers, below the Curie temperature, just as in the reference layers [47].

**Magnetotransport Properties**

Magnetotransport properties of conducting ferromagnets, such as magnetoresistance and anomalous and planar Hall effects, strongly depend on the strength of spin–orbit coupling in the material [31,60,61]. Typical dependences of Hall resistance $R_{xy}$ on the perpendicular magnetic field $B_\perp$ for the Hall-bars, prepared from the layers whose magnetic properties are presented in Figs. 3 and 4, measured at low temperatures 1.6 K and 4.2 K, are shown in Fig. 5 [56]. Microscopic image of the Hall-bar, showing its dimensions and crystallographic orientation, is presented in the inset.

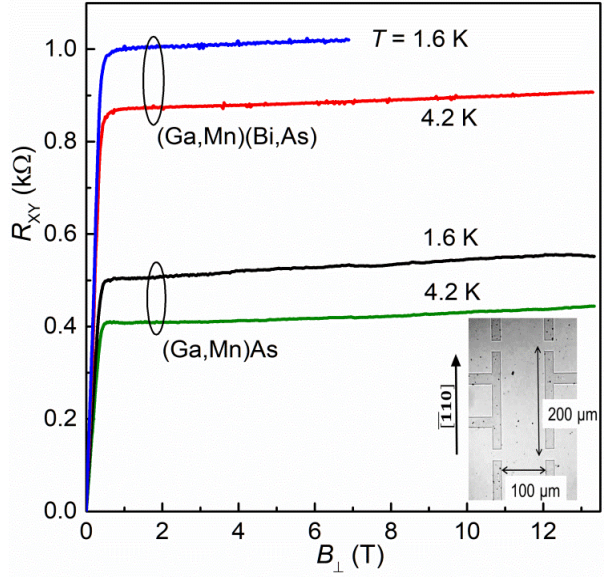

**Fig. 5** Hall resistance measured at temperatures of 1.6 K and 4.2 K for the Hall-bars of 10-nm thick (Ga,Mn)(Bi,As) and (Ga,Mn)As layers, with 6% Mn and 1% Bi contents, grown on GaAs, as a function of an external magnetic field $B_\perp$ perpendicular to the layer plane. Microscopic image of the Hall-bar and its geometry is shown in the inset, where the darker contrast corresponds to non-conducting areas etched to the substrate. After [56].

The Hall resistivity of conducting magnetic materials can be described by following equation [60]:

$$\rho_{xy} = R_H B_\perp + R_s M_\perp, \qquad (1)$$

where the first term represents the classical Hall effect, proportional to perpendicular magnetic field, which is commonly used to determine the type and concentration of free carriers in the material. The second term, called the anomalous Hall effect (AHE), originating from the spin–orbit interaction in the material, is proportional to the perpendicular component of its magnetization $M_\perp$ and dominates at low magnetic fields. Significantly larger magnitude of AHE for the (Ga,Mn)(Bi,As) Hall-bar than that for the reference one results from the enhanced SOC strength caused by the Bi incorporation in the DFS layer.

On the other hand, the classical Hall effect, predominant at high magnetic fields where the variation of AHE with a magnetic field is sufficiently small, displays similar magnitudes for both the layers. The hole concentrations, determined from the high-field results (of above 1 T), presented in Fig. 5, amount to about $2 \times 10^{20}$ cm$^{-3}$ for both the (Ga,Mn)(Bi,As) and (Ga,Mn)As layers at T = 4.2 K. Similar hole concentrations in other sets of the (Ga,Mn)(Bi,As) and reference (Ga,Mn)As layers have also been determined from our room-temperature Raman scattering spectroscopy measurements [35,62].

The longitudinal resistance $R_{xx}$ of the same Hall-bars, normalized to zero-field resistance, vs. perpendicular magnetic field, measured at temperatures of 1.6 K and 4.2 K is shown in Fig. 6 [56]. At low magnetic fields, $|B_\perp| < 0.4$ T, the results show a positive magnetoresistance (MR). It has been interpreted as a result of anisotropic magnetoresistance (AMR), the effect commonly appearing in conducting ferromagnetic materials, where their longitudinal resistivity $\rho_{xx}$ can be described by the following dependence on the angle $\theta$ between the electric current direction and the magnetization vector [63]:

$$\rho_{xx} = \rho_\perp + (\rho_\parallel - \rho_\perp)\cos^2\theta, \qquad (2)$$

where $\rho_\perp$ and $\rho_\parallel$ are the resistivities for the magnetization vector oriented perpendicular and parallel to the current direction, respectively.

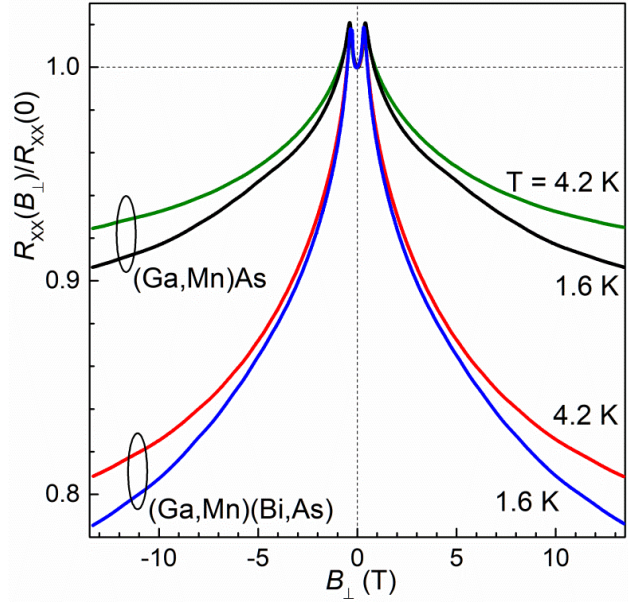

**Fig. 6** Relative longitudinal resistance measured for the same Hall-bars as in Fig. 5 at temperatures of 1.6 K and 4.2 K, while sweeping an external magnetic field $B_\perp$ perpendicular to the layer plane in opposite directions. After [56].

In contradiction to majority of metallic ferromagnets, where $\rho_\parallel$ is larger than $\rho_\perp$, in ferromagnetic (Ga,Mn)As layers $\rho_\parallel < \rho_\perp$, as revealed experimentally [22] and also explained theoretically [64]. Thus the positive MR, appearing at low magnetic field in Fig. 6, results from the rotation of the magnetization vector in the layer from the in-plane easy magnetization axis, at zero magnetic field, to the out-of-plane direction, at $B_\perp$ corresponding to the perpendicular anisotropy field, for which Equation (2) reaches the maximum value.

At higher magnetic fields, where the magnetization vector is aligned along the field, the $R_{xx}$ vs. $B_\perp$ dependence exhibits pronounced negative MR, whose magnitude increases with decreasing temperature and appears much larger for the Bi-contained layer. Such negative MR, commonly observed for the (Ga,Mn)As DFS layers at low temperatures, has been interpreted as a result of manifestation of the weak localization (WL) phenomenon [65–68]. WL results from the constructive quantum interference of two partial waves of a charge carrier moving diffusively along a closed trajectory in reverse directions, thus giving rise to the enhanced probability of backscattering and a positive contribution to electrical resistivity. A magnetic flux penetrating the closed trajectory causes a phase difference between the time-reversed interfering waves thus suppressing WL and leading to negative MR. On the other hand, the weak antilocalization (WAL) phenomenon, resulting from strong spin–orbit coupling in paramagnetic materials and leading to a positive MR at low magnetic fields, is quenched in ferromagnetic materials by the internal magnetic field [67,68].

To account quantitatively for the dependences of negative MR on magnetic field, shown in Fig. 6, we have adapted the theory of weak localization for two-dimensional ferromagnetic systems in perpendicular magnetic field developed by Dugaev et al. [67]. The theory considers WL quantum correction to conductivity taking into account the spin–orbit interaction, which manifests itself as spin–orbit scattering and is represented in the theoretical expressions by the spin–orbit scattering length $L_{so}$. By fitting the theoretical expression describing the dependence of this correction on magnetic field to our experimental results we have obtained a fine agreement between the theory and the experiment for both the investigated layers [56]. The $L_{so}$ values obtained from the fitting clearly demonstrate a considerable decrease in the $L_{so}$ value (from 140 nm in (Ga,Mn)As to 70 nm in (Ga,Mn)(Bi,As) at T = 4.2 K) as a result of 1% Bi incorporation in the (Ga,Mn)As layer. This decrease, by the factor of 2, gives another evidence for a significant impact of the enhanced SOC strength on magnetotransport phenomena in (Ga,Mn)As-based DFS layers.

Moreover, by taking into account the measured hole concentrations in the layers and their semi-classical Boltzmann resistivities obtained from the fitting we have calculated hole mobilities in the investigated layers [56]. A distinct decrease in the mobility caused by Bi addition to the (Ga,Mn)As layer (from 1.7 cm$^2$/Vs to 1.2 cm$^2$/Vs at T = 4.2 K) arises from the increased chemical disorder in the (Ga,Mn)(Bi,As) layers, cf. [69] and, possibly, the enhanced hole effective mass, as predicted by Pettinari et al. [70]. The relatively low mobility of holes, of the order of 1 cm$^2$/Vs, is generally observed in ferromagnetic (Ga,Mn)As layers and has been explained as due to a disordered character of the top of the valence band resulted from merging the host GaAs valence band with Mn-related impurity band in heavily Mn-doped layers; see, e.g., [16,71].

The planar Hall effect, appearing as spontaneous transverse voltage developing in response to longitudinal current flowing along a conducting ferromagnetic material in absence of applied magnetic field, is another manifestation of the spin–orbit interaction [72]. We have investigated the PHE resistivity employing the micro-Hall-bars of small dimensions, of 20 μm width and 50 μm distance between the voltage contacts, to ensure the presence of single magnetic domain inside the bar; cf., [73,74]. An example of such Hall-bar is presented in the inset in Fig. 7. The PHE resistivity in a ferromagnetic layer containing a single magnetic domain can be described by the expression [63]:

$$\rho_{xy} = \tfrac{1}{2}(\rho_\parallel - \rho_\perp)\sin 2\theta, \qquad (3)$$

An external in-plane magnetic field applied at an angle $\varphi$ with respect to the current direction (see inset in Fig. 7) allows of a change of the magnetization vector direction in the layer, thus changing the PHE resistivity according to Equation (3).

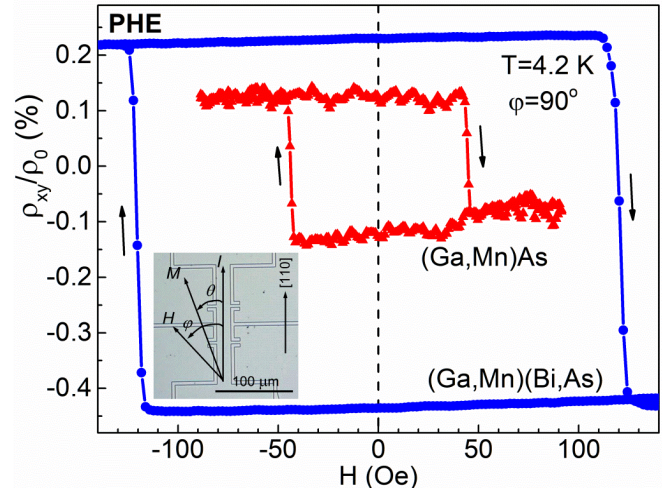

**Fig. 7** Normalized PHE resistivity for the (Ga,Mn)(Bi,As) and (Ga,Mn)As micro-Hall-bars measured at the temperature 4.2 K, while sweeping an in-plane magnetic field, perpendicular to the current ($\varphi = 90°$), in opposite directions, as indicated by the arrows. Inset shows microscopic image of the Hall-bar, tailored from the investigated (Ga,Mn)(Bi,As) layer along the [110] crystallographic direction, and configuration of the PHE measurements. $\theta$ and $\varphi$ denote angles between the current direction $I$ and the magnetization vector $M$ and applied in-plane magnetic field $H$, respectively. After [75].

Representative dependence of the PHE resistivity, normalized to zero-field longitudinal resistivity $\rho_0$, on magnetic field for micro-Hall-bars prepared from the 50-nm thick (Ga,Mn)(Bi,As) layer, with 6% Mn and 1% Bi contents, and the reference (Ga,Mn)As one is presented in Fig. 7 [75]. Here, the up and down sweep of the weak magnetic field applied along the [–110] crystallographic direction, perpendicular to the current, results in rotation of the magnetization vector in the Hall-bar between the easy magnetization axes along the [010] and [100] directions, respectively. The measured PHE resistivity displays the hysteresis loop dependence, similar to the magnetization hysteresis loop, described by Equation (3) where the angle $\theta$ changes between 45° and 315°. Accordingly, the PHE resistivity changes its value between $-\frac{1}{2}(\rho_\perp - \rho_\parallel)$ and $+\frac{1}{2}(\rho_\perp - \rho_\parallel)$. The difference between the high and low values of the hysteresis loops shown in Fig. 7 equals $(\rho_\perp - \rho_\parallel)/\rho_0$ for both the Hall-bars. Its value for the (Ga,Mn)(Bi,As) Hall-bar is by a factor of about 2.5 larger than that for the reference (Ga,Mn)As one as a result of the enhanced SOC strength. Similarly, the width of hysteresis loop, i.e., the coercive field, is larger by a factor of about 2.7 for the Bi-containing layer, in agreement with the magnetization coercive fields dependence for the hard in-plane magnetization axis. Note, however, that the SQUID magnetometry results, such as those shown in Fig. 4, have been obtained for multidomain layers of the area of about 20 mm$^2$.

Abrupt changes in the PHE resistivity at the coercive fields confirm the presence of single domains in hall-bars of both investigated layers. Earlier studies by means of a scanning SQUID microscope [73] and magneto-optical domain imaging [74] revealed, in (Ga,Mn)As layers with in-plane magnetization, large domains of a fraction of mm dimensions. Magnetization vector rotation in those layers, between the in-plane easy axes, resulted from the nucleation and propagation of a 90° domain wall arising in a small field range around the coercive field [74]. Our PHE results suggest similar behaviour in the (Ga,Mn)(Bi,As) layers.

Our very recent studies of the valence band structure performed for the 100-nm thick (Ga,Mn)(Bi,As) and reference (Ga,Mn)As layers, with 4% Mn and 0.3% Bi contents, by means of hard X-ray angle-resolved photoemission spectroscopy confirmed the Fermi level location within the valence band of the layers [62]. These findings are consistent with the valence band origin of the holes mediated ferromagnetic ordering in the layers. Moreover, a small atomic fraction of just 0.3% Bi content in the (Ga,Mn)As DFS results in the downward shift of both the light-hole and heavy-hole valence bands by about 150 meV with respect to the Fermi level and a significant increase in the spin–orbit split-off band separation from the valence band edge. Spectroscopic ellipsometry and modulation photoreflectance spectroscopy results confirm the Bi incorporation induced modifications in the valence bands of the layers [62].

The enhanced SOC strength caused by Bi incorporation into the (Ga,Mn)As DFS is advantageous for current-induced magnetization manipulation driven by the spin–orbit torque mechanism. We have investigated the SOT driven magnetization reversal in 15-nm thick (Ga,Mn)(Bi,As) layer, with 6% Mn and 1% Bi contents, grown on (In,Ga)As buffer under tensile misfit strain resulting in perpendicular magnetic anisotropy in the layer [76]. Perpendicular magnetic anisotropy is beneficial for low-power spintronic applications because of the capability of higher device density and higher thermal stability; see, e.g., [33,34,77]. In our experiments the magnetization reversal, triggered off by the injecting in-plane current pulses of either positive or negative sign, has been revealed by recording the anomalous Hall resistance under assistance of weak magnetic field parallel to the current. The current density threshold of about $5 \times 10^4$ A/cm$^2$ necessary for the reversal of magnetization direction at the temperature of 120 K in our experiments [76] is almost 6 times smaller than the one needed for the magnetization reversal in the research by Jiang et al. [78] performed for a tensile strained Bi-free (Ga,Mn)As layer with the same Mn content and even larger magnitude of the biaxial misfit strain. Moreover, this current density threshold is almost 3 orders of magnitude smaller than that in metallic systems composed of heavy-metal/ferromagnet heterostructures; c.f. [30,34]. It turns out also smaller than those recently demonstrated experimentally in other novel systems proposed for applications to SOT-driven memory and logic devices, like the ones containing magnetically doped topological insulators; cf. [79,80] and van der Waals-based two-dimensional transition metal dichalcogenide/ferromagnet heterostructures; see e.g. [81,82]. The requirement for assistance of weak in-plane magnetic field necessary for SOT driven magnetization reversal in our experiments, which is impractical for spintronic applications, can be overcome by using DFS layers of the wedge form displaying lateral structural asymmetry. Similar structures with high lateral structural asymmetry, resulting in the very strong Rashba SOC effect [28], have recently been proposed for various metallic systems designed for field-free SOT driven memory and logic applications [33,34,83].

**Conclusions**

Detailed research of the effect of bismuth incorporation into the (Ga,Mn)As dilute ferromagnetic semiconductor on its structural, magnetic and magnetotransport properties is encapsulated in this paper. It is demonstrated that (Ga,Mn)(Bi,As) epitaxial layers, containing up to 1% of large Bi atoms, with a high structural perfection and magnetic homogeneity, can be grown by precisely optimized molecular-beam epitaxy. The incorporation of small atomic fraction of heavy Bi atoms, substituting As atoms in the layers, gives rise to significant enhancement of the spin–orbit interaction, which strongly impacts on the magnetotransport phenomena in the layers, such as magnetoresistance and anomalous and planar Hall effects.

The low-temperature huge negative magnetoresistance has been explained as resulted from suppression of weak localization in magnetic field and quantitatively described by the WL theory for two-dimensional ferromagnets. Fitting this theory to the experimental data indicates a twofold decrease in the spin–orbit scattering length caused by 1% Bi addition to the (Ga,Mn)As layer. Moreover, the same amount of Bi in the layer results in a six-fold lowering the current density threshold necessary for the reversal of magnetization direction in the layer, driven by the spin–orbit torque mechanism, which is especially important for applications of this mechanism in the prospective, energy efficient, nonvolatile memory and logic elements. Relatively low values of ferromagnetic Curie temperature, of below 200 K as yet, of both the (Ga,Mn)As and (Ga,Mn)(Bi,As) DFS compounds make them not useful at present for applications in functional devices. Nonetheless, they may be advantageous for the research of novel spintronic functionalities making use of electrically manipulated ferromagnetism.

**Acknowledgments** The author would like to thank all the coauthors of the common papers summarized in this review for many years of fruitful collaboration and discussions.

**Conflict of interest** The author declares that he has no conflict of interest.

**References**

1. H. Ohno, A. Shen, F. Matsukura, A. Oiwa, A. Endo, S. Katsumoto, Y. Iye, (Ga,Mn)As: A new diluted magnetic semiconductor based on GaAs. Appl. Phys. Lett. 69, 363–365 (1996).
2. H. Ohno, Making nonmagnetic semiconductors ferromagnetic. Science 281, 951–955 (1998).
3. T. Dietl, H. Ohno, Dilute ferromagnetic semiconductors: Physics and spintronic structures. Rev. Mod. Phys. 86, 187–251 (2014).
4. T. Jungwirth, J. Wunderlich, V. Novak, K. Olejnik, B.L. Gallagher, R.P. Campion, K.W. Edmonds, A.W. Rushforth, A.J. Ferguson, P. Nemec, Spin-dependent phenomena and device concepts explored in (Ga,Mn)As. Rev. Mod. Phys. 86, 855–896 (2014).
5. D. Chiba, Y. Nishitani, F. Matsukura, H. Ohno, Properties of $Ga_{1-x}Mn_xAs$ with high Mn composition (x>0.1). Appl. Phys. Lett. 90, 122503 (2007).
6. S. Mack, R.C. Myers, T. Heron, A.C. Gossard, D.D. Awschalom, Stoichiometric growth of high Curie temperature heavily alloyed GaMnAs. Appl. Phys. Lett. 92, 192502 (2008).
7. P. Nemec, V. Novak, N. Tesarova, E. Rozkotova, H. Reichlova, D. Butkovicova, F. Trojanek, K. Olejnik, P. Maly, R.P. Campion, B.L. Gallagher, J. Sinova, T. Jungwirth, The essential role of carefully optimized synthesis for elucidating intrinsic material properties of (Ga,Mn)As. Nature Commun. 4, 1422 (2013).
8. T. Dietl, H. Ohno, F. Matsukura, J. Cibert, D. Ferrand, Zener model description of ferromagnetism in zinc-blende magnetic semiconductors. Science 287, 1019–1022 (2000).

9. T. Dietl, H. Ohno, F. Matsukura, Hole-mediated ferromagnetism in tetrahedrally coordinated semiconductors. Phys. Rev. B 63, 195205 (2001).
10. M. Berciu, R.N. Bhatt, Mean-field approach to ferromagnetism in (III,Mn)V diluted magnetic semiconductors at low carrier densities. Phys. Rev. B 69, 045202 (2004).
11. Y. Zhang, S. Das Sarma, Temperature and magnetization-dependent band gap renormalization and optical many-body effects in diluted magnetic semiconductors. Phys. Rev. B 72, 125303 (2005).
12. B.L. Sheu, R.C. Myers, J.-M. Tang, N. Samarth, D.D. Awschalom, P. Schiffer, M.E. Flatté, Onset of ferromagnetism in low-doped $Ga_{1-x}Mn_xAs$. Phys. Rev. Lett. 99, 227205 (2007).
13. K. Alberi, K.M. Yu, P.R. Stone, O.D. Dubon, W. Walukiewicz, T. Wojtowicz, X. Liu, J.K. Furdyna, Formation of Mn-derived impurity band in III-Mn-V alloys by valence band anticrossing. Phys. Rev. B 78, 075201 (2008).
14. J.B. Fujii, R. Salles, M. Sperl, S. Ueda, M. Kobata, K. Kobayashi, Y. Yamashita, P. Torelli, M. Utz, C.S. Fadley, A.X. Gray, J. Braun, H. Ebert, I. Di Marco, O. Eriksson, P. Thunström, G.H. Fecher, H. Stryhanyuk, E. Ikenaga, J. Minár, C.H. Back, G. van der Laan, G. Panaccione, Identifying the electronic character and role of the Mn states in the valence band of (Ga,Mn)As. Phys. Rev. Lett. 111, 097201 (2013).
15. O. Yastrubchak, J. Zuk, H. Krzyżanowska, J.Z. Domagala, T. Andrearczyk, J. Sadowski, T. Wosinski, Photorefectance study of the fundamental optical properties of (Ga,Mn)As epitaxial flms. Phys. Rev. B 83, 245201 (2011).
16. O. Yastrubchak, J. Sadowski, H. Krzyżanowska, L. Gluba, J. Żuk, J.Z. Domagala, T. Andrearczyk, T. Wosinski, Electronic- and band-structure evolution in low-doped (Ga,Mn)As. J. Appl. Phys. 115, 053710 (2013).
17. L. Gluba, O. Yastrubchak, J.Z. Domagala, R. Jakiela, T. Andrearczyk, J. Żuk, T. Wosinski, J. Sadowski, M. Sawicki, Band structure evolution and the origin of magnetism in (Ga,Mn)As: from paramagnetic through superparamagnetic to ferromagnetic phase. Phys. Rev. B 97, 115201 (2018).
18. M. Sawicki, F. Matsukura, A. Idziaszek, T. Dietl, G.M. Schott, C. Ruester, C. Gould, G. Karczewski, G. Schmidt, L.W. Molenkamp, Temperature dependent magnetic anisotropy in (Ga,Mn)As layers. Phys. Rev. B 70, 245325 (2004).
19. S. Hümpfner, K. Pappert, J. Wenisch, K. Brunner, C. Gould, G. Schmidt, L.W. Molenkamp, M. Sawicki, T. Dietl, Lithographic engineering of anisotropies in (Ga,Mn)As. Appl. Phys. Lett. 90, 102102 (2007).
20. J. Wunderlich, A.C. Irvine, J. Zemen, V. Holý, A.W. Rushforth, E. De Ranieri, U. Rana, K. Výborný, J. Sinova, C.T. Foxon, R.P. Campion, D.A. Williams, B.L. Gallagher, T. Jungwirth, Local control of magnetocrystalline anisotropy in (Ga,Mn)As microdevices: Demonstration in current-induced switching. Phys. Rev. B 76, 054424 (2007).
21. J. Wenisch, C. Gould, L. Ebel, J. Storz, K. Pappert, M.J. Schmidt, C. Kumpf, G. Schmidt, K. Brunner, L.W. Molenkamp, Control of magnetic anisotropy in (Ga,Mn)As by lithography-induced strain relaxation. Phys. Rev. Lett. 99, 077201 (2007).
22. H.X. Tang, R.K. Kawakami, D.D. Awschalom, M.L. Roukes, Giant planar Hall effect in epitaxial (Ga,Mn)As devices. Phys. Rev. Lett. 90, 107201 (2003).
23. S. Das Sarma, Ferromagnetic semiconductors: A giant appears in spintronics. Nat. Mater. 2, 292–294 (2003).

24. S. Lee, D.Y. Shin, S.J. Chung, X. Liu, J.K. Furdyna, Tunable quaternary states in ferromagnetic semiconductor GaMnAs single layer for memory devices. Appl. Phys. Lett. 90, 152113 (2007).
25. T. Wosinski, T. Andrearczyk, T. Figielski, J. Wrobel, J. Sadowski, Domain-wall controlled (Ga,Mn)As nanostructures for spintronic applications. Physica E 51, 128–134 (2013).
26. A. Chernyshov, M. Overby, X. Liu, J.K. Furdyna, Y. Lyanda-Geller, L.P. Rokhinson, Evidence for reversible control of magnetization in a ferromagnetic material by means of spin–orbit magnetic field. Nature Physics 5, 656–659 (2009).
27. G. Dresselhaus, Spin–orbit coupling effects in zinc blende structures. Phys. Rev. 100, 580–586 (1955).
28. Y.A. Bychkov, E.I. Rashba, Properties of a 2D electron gas with lifted spectral degeneracy. JETP Lett. 39, 78–81 (1984).
29. D. Fang, H. Kurebayashi, J. Wunderlich, K. Vyborny, L.P. Zarbo, R.P. Campion, A. Casiraghi, B.L. Gallagher, T. Jungwirth, and A.J. Ferguson, Spin-orbit driven ferromagnetic resonance. Nature Nanotech. 6, 413–417 (2011).
30. I.M. Miron, K. Garello, G. Gaudin, P.-J. Zermatten, M.V. Costache, S. Auffret, S. Bandiera, B. Rodmacq, A. Schuhl, P. Gambardella, Perpendicular switching of a single ferromagnetic layer induced by in-plane current injection. Nature 476, 189–194 (2011).
31. A. Manchon, J. Zelezny, I.M. Miron, T. Jungwirth, J. Sinova, A. Thiaville, K. Garello, P. Gambardella, Current-induced spin–orbit torques in ferromagnetic and antiferromagnetic systems. Rev. Mod. Phys. 91, 035004 (2019).
32. R. Ramaswany, J.M. Lee, K. Cai, H. Yang, Recent advances in spin–orbit torques: Moving towards device applications. Appl. Phys. Rev. 5, 031107 (2018).
33. V. Lopez-Dominguez, Y. Shao, P.K. Amiri, Perspectives on field-free spin–orbit torque devices for memory and computing applications. J. Appl. Phys. 133, 040902 (2023).
34. Q. Yang, D. Han, S. Zhao, J. Kang, F. Wang, S.-C. Lee, J. Lei, K.-J. Lee, B.-G. Park, H. Yang, Field-free spin–orbit torque switching in ferromagnetic trilayers at sub-ns timescales. Nature Commun. 15, 1814 (2024).
35. O. Yastrubchak, J. Sadowski, L. Gluba, J.Z. Domagala, M. Rawski, J. Żuk, M. Kulik, T. Andrearczyk, T. Wosinski, Ferromagnetism and the electronic band structure in (Ga,Mn)(Bi,As) epitaxial layers. Appl. Phys. Lett. 105, 072402 (2014).
36. K. Levchenko, T. Andrearczyk, J.Z. Domagala, J. Sadowski, L. Kowalczyk, M. Szot, R. Kuna, T. Figielski, T. Wosinski, Novel quaternary dilute magnetic semiconductor (Ga,Mn)(Bi,As): Magnetic and magneto-transport investigations. J. Supercond. Nov. Magn. 30, 825–829 (2017).
37. B. Fluegel, S. Francoeur, A. Mascarenhas, S. Tixier, E.C. Young, T. Tiedje, Giant spin–orbit bowing in GaAs$_{1-x}$Bi$_x$. Phys. Rev. Lett. 97, 067205 (2006).
38. K. Alberi, J. Wu, W. Walukiewicz, K.M. Yu, O.D. Dubon, S.P. Watkins, C.X. Wang, X. Liu, Y.-J. Cho, J. Furdyna, Valence band anticrossing in mismatched III-V semiconductor alloys. Phys. Rev. B 75, 045203 (2007).
39. M. Usman, C.A. Broderick, Z. Batool, K. Hild, T.J.C. Hosea, S.J. Sweeney, E.P. O'Reilly, Impact of alloy disorder on the band structure of compressively strained GaBi$_x$As$_{1-x}$. Phys. Rev. B 87, 115104 (2013).
40. T.F. Kuech, S.E. Babcock, L. Mawst, Growth far from equilibrium: Examples from III-V semiconductors. Appl. Phys. Rev. 3, 040801 (2016).

41. L. Wang, L. Zhang, L. Yue, D. Liang, X. Chen, Y. Li, P. Lu, J. Shao, S. Wang, Novel dilute bismide, epitaxy, physical properties and device application. Crystals 7, 63 (2017).
42. Ł. Gelczuk, J. Kopaczek, T.B.O. Rockett, R.D. Richards, R. Kudrawiec, Deep-level defects in n-type GaAsBi alloys grown by molecular beam epitaxy at low temperature and their influence on optical properties. Sci. Rep. 7, 12824 (2017).
43. M. Fregolent, M. Buffolo, C. De Santi, S. Hasegawa, J. Matsumura, H. Nishinaka, M. Yoshimoto, G. Meneghesso, E. Zanoni, M. Maneghini, Deep levels and carrier capture kinetics in n-GaAsBi alloys investigated by deep level transient spectroscopy. J. Phys. D: Appl. Phys. 54, 345109 (2021).
44. Ł. Gelczuk, J. Kopaczek, D. Pucicki, T.B.O. Rockett, R.D. Richards, R. Kudrawiec, Effects of rapid thermal annealing on deep-level defects and optical properties of n-type GaAsBi alloys grown by molecular beam epitaxy at low temperature. Mater. Sci. Semicond. Process. 169, 107888 (2024).
45. C. Himwas, A. Soison, S. Kijamnajsuk, T. Wongpinij, C. Euraksakul, S. Panyakeow, S. Kanjanachuchai, GaAsPBi epitaxial layer grown by molecular beam epitaxy. Semicond. Sci. Technol. 35, 095009 (2020).
46. M. Aydin, S.N. Yilmaz, A. Erol, J. Bork, J. Zide, O. Donmez, Electron energy relaxation mechanism in n-type $In_xGa_{1-x}As_{1-y}Bi_y$ alloys under electric and magnetic fields. Phys. Scr. 99, 105909 (2024).
47. K. Levchenko, T. Prokscha, J. Sadowski, I. Radelytskyi, R. Jakiela, M. Trzyna, T. Andrearczyk, T. Figielski, T. Wosinski, Evidence for the homogeneous ferromagnetic phase in (Ga,Mn)(Bi,As) epitaxial layers from muon spin relaxation spectroscopy. Sci. Rep. 9, 3394 (2019).
48. J. Sadowski, J.Z. Domagała, J. Bąk-Misiuk, S. Koleśnik, M. Sawicki, K. Świątek, J. Kanski, L. Ilver, V. Ström, Structural and magnetic properties of molecular beam epitaxy grown GaMnAs layers. J. Vac. Sci. Technol. B 18, 1697–1700 (2000).
49. K.W. Edmonds, P. Bogusławski, K.Y. Wang, R.P. Campion, S.V. Novikov, N.R.S. Farley, B.L. Gallagher, C.T. Foxon, M. Sawicki, T. Dietl, M. Boungiorno Nardelli, J. Bernholc, Mn interstitial diffusion in (Ga,Mn)As. Phys. Rev. Lett. 92, 037201 (2004).
50. I. Kuryliszyn-Kudelska, J.Z. Domagała, T. Wojtowicz, X. Liu, E. Łusakowska, W. Dobrowolski, J.K. Furdyna, Effect of Mn interstitials on the lattice parameter of $Ga_{1-x}Mn_xAs$. J. Appl. Phys. 95, 603–608 (2004).
51. J. Puustinen, M. Wu, E. Luna, A. Schramm, P. Laukkanen, M. Laitinen, T. Sajavaara, M. Guina, Variation of lattice constant and cluster formation in GaAsBi. J. Appl. Phys. 114, 243504 (2013).
52. T. Prokscha, E. Morenzoni, K. Deiters, F. Foroughi, D. George, R. Kobler, A. Suter, V. Vrankovic, The new beam at PSI: A hybrid-type large acceptance channel for the generation of a high intensity surface-muon beam. Nucl. Instrum. Methods Phys. Res. A 595, 317 (2008).
53. C. Schlueter, A. Gloskovskii, K. Ederer, S. Piec, M. Sing, R. Claessen, C. Wiemann, C.-M. Schneider, K. Medjanik, G. Schönhense, P. Amann, A. Nilsson, W. Drube, The New Dedicated HAXPES Beamline P22 at PETRAIII". in *AIP Conference Proceedings* No. 2054, p. 040010 (2019).
54. T. Andrearczyk, K. Levchenko, J. Sadowski, J.Z. Domagala, A. Kaleta, P. Dłużewski, J. Wróbel, T. Figielski, T. Wosinski, Structural quality and magnetotransport properties of epitaxial layers of the (Ga,Mn)(Bi,As) dilute magnetic semiconductor. Materials 13, 5507 (2020).

55. O. Yastrubchak, T. Wosiński, J.Z. Domagała, E. Łusakowska, T. Figielski, B. Pécz, A.L. Tóth, Misfit strain anisotropy in partially relaxed lattice-mismatched InGaAs/GaAs heterostructures. J. Phys.: Condens. Matter 16, S1–S8 (2004).
56. T. Andrearczyk, K. Levchenko, J. Sadowski, K. Gas, A. Avdonin, J. Wróbel, T. Figielski, M. Sawicki, T. Wosinski, Impact of bismuth incorporation into (Ga,Mn)As dilute ferromagnetic semiconductor on its magnetic properties and magnetoresistance. Materials 16, 788 (2023).
57. K.-Y. Wang, M. Sawicki, K.W. Edmonds, R.P. Campion, S. Maat, C.T. Foxon, B.L. Gallagher, T. Dietl, Spin Reorientation Transition in Single-Domain (Ga,Mn)As. Phys. Rev. Lett. 95, 217204 (2005).
58. N. Tataryn, L. Gluba, O. Yastrubchak, J. Sadowski, T. Andrearczyk, S. Mamykin, M. Sawicki, T. Wosinski, Valence band dispersion in Bi doped (Ga,Mn)As epitaxial layers. IEEE Trans. Magn. 59, 4100405 (2023).
59. M. Birowska, C. Sliwa, J.A. Majewski, T. Dietl, Origin of bulk uniaxial anisotropy in zinc-blende dilute magnetic semiconductors. Phys. Rev. Lett. 108, 237203 (2012).
60. N. Nagaosa, J. Sinova, S. Onoda, A.H. MacDonald, N.P. Ong, Anomalous Hall effect. Rev. Mod. Phys. 82, 1539–1592 (2010).
61. J. Sinova, S.O. Valenzuela, J. Wunderlich, C.H. Back, T. Jungwirth, Spin Hall effects. Rev. Mod. Phys. 87, 1213–1259 (2015).
62. O. Yastrubchak, N. Tataryn, L. Gluba, S. Mamykin, J. Sadowski, T. Andrearczyk, J.Z. Domagala, O. Kondratenko, V. Romanyuk, O. Fedchenko, Y. Lytvynenko, O. Tkach, D. Vasilyev, S. Babenkov, K. Medjanik, K. Gas, M. Sawicki, T. Wosinski, G. Schönhense, H.-J. Elmers, Influence of Bi doping on the electronic structure of (Ga,Mn)As epitaxial layers. Sci. Rep. 13, 17278 (2023).
63. T.R. McGuire, R.I. Potter, Anisotropic magnetoresistance in ferromagnetic $3d$ alloys. IEEE Trans. Magn. 11, 1018–1038 (1975).
64. T. Jungwirth, J. Sinova, K.Y. Wang, K.W. Edmonds, R.P. Campion, B.L. Gallagher, C.T. Foxon, Q. Niu, A.H. MacDonald, Dc-transport properties of ferromagnetic (Ga,Mn)As semiconductors. Appl. Phys. Lett. 83, 320-322 (2003).
65. F. Matsukura, M. Sawicki, T. Dietl, D. Chiba, H. Ohno, Magnetotranport properties of metallic (Ga,Mn)As films with compressive and tensile strain. Physica E 21, 1032–1036 (2004).
66. T. Dietl, Interplay between carrier localization and magnetism in diluted magnetic and ferromagnetic semiconductors. J. Phys. Soc. Jpn. 77, 031005 (2008).
67. V.K. Dugaev, P. Bruno, J. Barnaś, Weak localization in ferromagnets with spin–orbit interaction. Phys. Rev. B 64, 144423 (2001).
68. I. Garate, J. Sinova, T. Jungwirth, A.H. MacDonald, Theory of weak localization in ferromagnetic (Ga,Mn)As. Phys. Rev. B 79, 155207 (2009).
69. R.N. Kini, A.J. Ptak, B. Fluegel, R. France, R.C. Reedy, A. Mascarenhas, Effect of Bi alloying oh the hole transport in the dilute bismide alloy GaAs$_{1-x}$Bi$_x$. Phys. Rev. B 83, 075307 (2011).
70. G. Pettinari, M. Capizzi, A. Polimeni, Carrier masses and band-gap temperature sensitivity in Ga(AsBi) alloys. Semicond. Sci. Technol. 30, 094002 (2015).
71. T. Jungwirth, J. Sinova, A.H. MacDonald, B.L. Gallagher, V. Novák, K.W. Edmonds, A.W. Rushforth, R.P. Campion, C.T. Foxon, L. Eaves, E. Olejník, J. Masek, S.-R. Eric Yang, J. Wunderlich, C. Gould, L.W. Molenkamp, T. Dietl, H. Ohno, Character of states near the Fermi level in (Ga,Mn)As: Impurity to valence band crossover. Phys. Rev. B 76, 125206 (2007).

72. C. Goldberg, R.E. Davis, New galvanomagnetic effect. Phys. Rev. 94, 1121–1125 (1954).
73. T. Fukumura, T. Shono, K. Inaba, T. Hasegawa, H. Koinuma, F. Matsukura, H. Ohno, Magnetic domain structure of a ferromagnetic semiconductor (Ga,Mn)As observed with scanning probe microscopes. Physica E 10, 135-138 (2001).
74. U. Welp, V.K. Vlasko-Vlasov, X. Liu, J.K. Furdyna, T. Wojtowicz, Magnetic domain structure and magnetic anisotropy in $Ga_{1-x}Mn_xAs$. Phys. Rev. Lett. 90, 167206 (2003).
75. T. Andrearczyk, J. Sadowski, J. Wróbel, T. Figielski, T. Wosinski, Tunable planar Hall effect in (Ga,Mn)(Bi,As) epitaxial layers. Materials 14, 4483 (2021).
76. T. Andrearczyk, J. Sadowski, K. Dybko, T. Figielski, T. Wosinski, Current-induced magnetization reversal in (Ga,Mn)(Bi,As) epitaxial layer with perpendicular magnetic anisotropy. Appl. Phys. Lett. 121, 242401 (2022).
77. R.A. Duine, K.J. Lee, S.S.P. Parkin, M.D. Stiles, Synthetic antiferromagnetic spintronics. Nat. Phys. 14, 217–219 (2018).
78. M. Jiang, H. Asahara, S. Sato, T. Kanaki, H. Yamasaki, S. Ohya, M. Tanaka, Efficient full spin-orbit torque switching in a single layer of a perpendicularly magnetized single-crystalline ferromagnet. Nat. Commun. 10, 2590 (2019).
79. Y. Fan, P. Upadhyaya, X. Kou, M. Lang, S. Takei, Z. Wang, J. Tang, L. He, L.-T. Chang, M. Montazeri, G. Yu, W. Jiang, T. Nie, R.N. Schwartz, Y. Tserkovnyak, K.L. Wang, Magnetization switching through giant spin–orbit torque in magnetically doped topological insulator heterostructure. Nature Mater. 13, 699–704 (2014).
80. M. Mogi, K. Yasuda, R. Fujimura, R. Yoshimi, N. Ogawa, A. Tsukazaki, M. Kawamura, K.S. Takahashi, M. Kawasaki, Y. Tokura, Current-induced switching of proximity-induced ferromagnetic surface states in a topological insulator. Nat. Commun. 12, 1404 (2021).
81. Y. Ou, W. Yanez, R. Xiao, M. Stanley, S. Ghosh, B. Zheng, W. Jiang, Y.-S. Huang, T. Pillsbury, A. Richardella, C. Liu, T. Low, V.H. Crespi, K.A. Mkhoyan, N. Samarth, $ZrTe_2$/ $CrTe_2$: an epitaxial van der Waals platform for spintronics. Nat. Commun. 13, 2972 (2022).
82. I.-H. Kao, R. Muzzio, H. Zhang, M. Zhu, J. Gobbo, S. Yuan, D. Weber, R. Rao, J. Li, J.H. Edgar, J.E. Goldberger, J. Yan, D.G. Mandrus, J. Hwang, R. Cheng, J. Katoch, S. Singh, Deterministic switching of a perpendicularly polarized magnet using unconventional spin–orbit torques in $WTe_2$. Nature Mater. 21, 1029–1034 (2022).
83. B. Cui, Z. Zhu, C. Wu, X. Guo, Z. Nie, H. Wu, T. Guo, P. Chen, D. Zheng, T. Yu, L. Xi, Z. Zeng, S. Liang, G. Zhang, G. Yu, K.L. Wang, Comprehensive study of the current-induced spin–orbit torque perpendicular effective field in asymmetric multilayers. Nanomaterials 12, 1887 (2022).